\documentclass[prd,aps,preprint,showpacs,nofootinbib,superscriptaddress,tightenlines]{revtex4-1}
\usepackage{amssymb,amsmath}
\usepackage{hyperref}
\usepackage[capitalise]{cleveref}
\usepackage{graphicx,subfigure}

\begin{document}
\title{\textbf{The BMS Equation and $c\bar{c}$ Production; A Comparison of the BMS and BK Equations}}
\author{Giuseppe Marchesini}
\affiliation{Dipartimento di Fisica, Universit\`a di Milano-Bicocca and INFN, Sezione di Milano-Bicocca, Italy}
\email{Giuseppe.Marchesini@mib.infn.it}
\author{A.H.~Mueller}
\affiliation{Department of Physics, Columbia University, New York, NY, USA}
\email{amh@phys.columbia.edu}

\begin{abstract}
We introduce two processes where the BMS equation appears in a context quite different from the original context of non-global jet observables. We note the strong similarities of the BMS equation to the BK and FKPP equations and argue that these, essentially identical equations, can be viewed either in terms of the probability, or amplitude, of something not happening or in terms of the nonlinear terms setting unitarity limits. Mostly analytic solutions are given for (i) the probability that no $c\bar{c}$ pairs be produced in a jet decay and (ii) the probability that no-$c\bar{c}$ pairs be produced in a high energy dipole nucleus scattering. Both these processes obey BMS equations, albeit with very different kernels. 
\end{abstract}

\maketitle

\section{Introduction}
The purpose of this paper is to examine the origins and properties of the Banfi, Marchesini, Smye (BMS) equation~\cite{Banfi:2002hw} and to compare it with the Balitsky-Kovchegov (BK)~\cite{Balitsky:1995ub,Kovchegov:1999yj} equation with which it has a strong resemblance. The BMS equation arose in the study of ``non-global'' observables in jet physics~\cite{Dasgupta:2001sh}, in particular it describes the probability that in an $e^{+}e^{-}$ annihilation at energy $E$ less than $E_{\mathrm{out}}$ energy flows into a fixed angular region away from the jet-axes~\cite{Banfi:2002hw,Dasgupta:2001sh}. On the other hand the BK equation generalizes the linear Balitsky, Fadin, Kuraev, Lipatov (BFKL)~\cite{Kuraev:1977fs,Balitsky:1978ic} equation adding a nonlinear term which imposes unitarity at high energy when the scattering gets strong. At first sight these two equations, BK and BMS, would seem to be describing different phenomena~\cite{Marchesini:2003nh,Marchesini:2004ne,Hatta:2008st}. We shall argue below that this is not the case and that the physics origins of the two are very closely related when one views the equations in a particular way.

Our first object is to argue that the BMS equation is not necessarily tied to jet physics~\cite{Caron-Huot:2015bja,Neill:2015nya} and that even when used in jet physics it does not necessarily have to be tied to the details of where decay products go. To that end in \cref{sec:prob} we begin by defining an observable, $G(Q)$, the probability that a jet of virtuality $Q$ have no $c\bar{c}$ pairs in its decay products. We note in \cref{eq:qgq} that $G$ obeys a BMS equation which we are able to solve analytically in the region $G=1-P$ near one and in the region $G$ near 0. In the region of moderate $Q$ where $P$ is small, $P$ increases according to the usual, angular ordered~\cite{Mueller:1981ex,Ermolaev:1981cm,Bassetto:1982ma}, formula for jet multiplicity growth~\cite{Mueller:1981ex,Bassetto:1982ma,khoze} while in the region where $G$ is small the $Q$-dependence is given by a Levin-Tuchin type of expression~\cite{Levin:1999mw,Iancu:2003uh}. We have been able to get analytic answers when $P$ or $G$ is small both using a fixed coupling and using a running coupling.

So why does a nonlinear BMS equation emerge for this observable? We believe that one must have two conditions for a BMS or BK equation to emerge. First, there must be a stochastic branching of one object to go to two objects. In the example of \cref{sec:prob} the stochastic branching is $g\rightarrow gg$. Secondly, there should be a probability of something \underline{not} happening. In the example of \cref{sec:prob} we evaluate the probability that \underline{no} $c\bar{c}$ pairs be produced. When $g\rightarrow g_{1}g_{2}$ the probability that no $c\bar{c}$ pairs be produced becomes the probability that $g_{1}$ not have $c\bar{c}$ pairs in its decay products times the probability that $g_{2}$ not have $c\bar{c}$ decay products. This is the nonlinear term on the right hand side of \cref{eq:qgq}.

In order to understand these conditions better in \cref{sec:branching} we review a classic result in statistical physics~\cite{McKean,Bramson,Brunet,Munier:2014bba}. If one has a branching (one particle going to two particles) random walk in one spatial dimension, the $x$-axis, starting with a single particle at the origin then the probability that \underline{no} particle reach a definite position $x$ at time $t$ is given by the solution to the Fisher; Kolmogorov, Petrovsky, Piscounov (FKPP) equation~\cite{Fisher,Kolmogorov}. (See \cref{eq:fkpp} below.) The argument for this equation is almost identical to that given in \cref{sec:prob}. However, there is a difference between \cref{eq:qgq,eq:fkpp}. For \cref{eq:fkpp} the initial condition is given by \cref{eq:condition}. That is at $x<0$ the initial condition is at the stable fixed point of \cref{eq:fkpp} while for $x>0$ it is at the unstable fixed point. \cref{eq:qgq} does not have the analog of the variable $x$ so it is not possible to move from one fixed point to another. The final term on the right hand side of \cref{eq:qgq} eliminates $G=1$ as a fixed point so that the natural initial condition is just $G=1$ at $Q=2M$, with $M$ the mass of the heavy quark. The term $-\alpha dG(Q)$ drives $G$ away from what would be the unstable fixed point in the absence of the $\alpha d$ term. $\alpha d$ is evaluated in \cref{app:a}.

The BK, or FKPP, equation has two fixed points with the initial condition in some regions at one of the fixed points and in other regions at the other fixed point. The BMS equation has one less variable and cannot go between the two fixed points. So there must be another term in the equation, the final term on the right hand side of \cref{eq:qgq} which allows the initial condition to be $G=1$. In the BMS equation there are always two channels, $g\rightarrow gg$ or $g\rightarrow c\bar{c}$ in the case at hand, with the $-\alpha dG$ term in \cref{eq:qgq} representing the loss in probability that no $c\bar{c}$ pairs be produced.

The BK equation, reviewed in \cref{sec:bk} is in the same universality class as the FKPP equation~\cite{Munier:2014bba,Munier:2003sj}. If one writes this equation in terms of the $S$-matrix, as in \cref{eq:bkS}. The interpretation is the same as for the FKPP equation. $S$ is the amplitude for no inelastic interaction happening when a dipole of size $x_{01}=|\underline{x}_{0}-\underline{x}_{1}|$ passes through a large nucleus. The initial condition goes from the unstable fixed point, when $x_{01}<1/Q_{s}$, to the stable fixed point, when $x_{01}>1/Q_{s}$. For the BK equation to be more than a mean field approximation a large nuclear target is necessary~\cite{Kovchegov:1999yj}. The BK equation properly treats the fluctuations (stochasticity) of the projectile, but not of the target. Thus it is necessary, in the case of scattering, to choose a target which is not stochastic. In the BMS equation of \cref{sec:prob} there is no target so this issue does not arise, however, in our example in \cref{sec:no} the BMS equation will only be correct, beyond a mean field approximation, for a large nuclear target.

In \cref{sec:no}. $G$ stands for the probability that no $c\bar{c}$ pairs be produced in a high energy dipole-nucleus collisions. In order to make the problem solvable near $G=0$ and near $G=1$ we limit the rapidity region $\Delta y$ where there are to be no $c\bar{c}$ pairs produced to be near the rapidity of the nucleus but with rapidity high enough to make the coherence length of the $c\bar{c}$ pairs larger than the nuclear diameter. The simplicity of this choice is that the saturation momentum, $Q_{s}$, in \cref{eq:Nv} can be taken to be rapidity independent and equal to the McLerran-Venugopalan saturation momentum. When $G$ is near one we find a BFKL growth of $1-G$ while when $G$ is near zero a Levin-Tuchin form emerges.

Our purpose here is mainly conceptual though both of the nonjet BMS processes that we have discussed here could have phenomenological interest. In particular if one were to choose $\Delta y$ in \cref{sec:no} to be the whole LHC rapidity interval one should get into the nonlinear regime of \cref{eq:bms}. However, because of the rapidity dependence of $Q_{s}$ the resulting equation is not exactly solvable, even in the $G\simeq 1$ region and so numerical solutions would be necessary.

Finally, a comment on the way we have viewed the BMS and BK equations. We have taken the view that both BMS and BK can be viewed as the probability, or amplitude, of something not happening. This is very natural for BMS examples, but perhaps less natural for the BK equation. For the BK equation, \cref{eq:bk}, the nonlinear term cuts down the rate of growth of $T$ so as to stay below the unitarity limit. One can also view the BMS equation as a unitarity imposing equation. If one writes the BMS equation \cref{eq:qgq} in terms of $P$ (The linear part of this equation is given in \cref{eq:qpq}.) then the nonlinear terms in $P$ require that $P$, the probability of at least one $c\bar{c}$ appearing in the decay, remains below one. This is indeed a unitarity condition just like the BK equation for $T$ given in \cref{eq:bk} where $T=1-S$ is the scattering amplitude.

\section{The probability of no $c\bar{c}$ pairs in a gluon jet\label{sec:prob}}
In this section we set up, and approximately solve, the equation for the probability of a gluon jet, of virtuality $Q$, not having any charm-anticharm pairs in its decay products. When $Q$ is not too much greater than the charm threshold, $2M$, we expect that probability to be near one while for extremely large values of $Q$ we expect the probability to be near zero. As will be shown below we are able to analytically solve for the $Q$-dependence of this probability when it is near one or when it is small, and we will give an equation for the probability for any value of $Q$.

\subsection{Fixed coupling evaluation}
\begin{figure}[h]
  \centering
  \includegraphics[width=13cm]{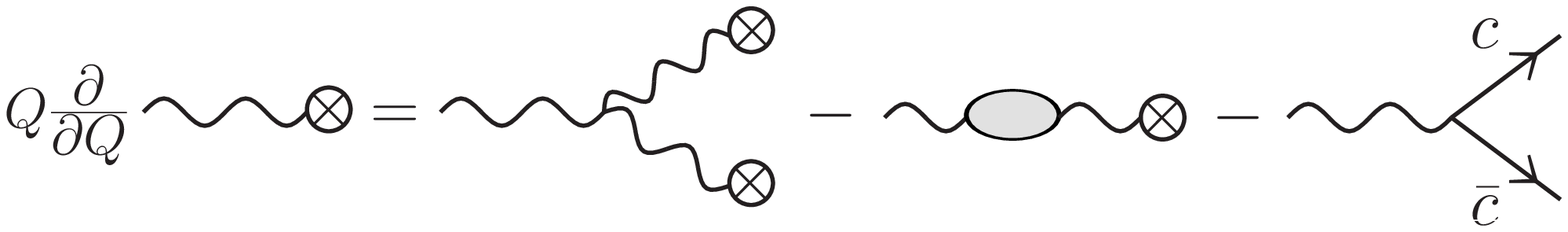}
\caption{}
\label{fig:splitting}
\end{figure}
Consider a gluon of virtuality $Q$ decaying either into two gluons or into a $c\bar{c}$ pair as illustrated in \cref{fig:splitting}. The process obeys the equation
\begin{equation}
  Q\frac{\partial}{\partial Q}G(Q)=\frac{2\alpha N_{c}}{\pi}\int^{1}_{0}\frac{dx}{x}\big[G(xQ)G\big((1-x)Q\big)-G(Q)\big]-\alpha dG(Q)
\label{eq:qgq}
\end{equation}
with $G(Q)$ the probability that no $c\bar{c}$ pairs have appeared in the branching up to $Q$, and with the initial condition $G(Q)=1$ for $Q=2M$. ($Q$ is given by the energy of the gluon, $E$, times the opening angle allowed at which it has been produced, $\Theta$, so that $Q=E\Theta$ is the natural evolution variable in an angular ordered jet cascade.) The three terms on the right hand side of \cref{eq:qgq} correspond to the three terms on the right hand side in \cref{fig:splitting}. At the moment we use a fixed coupling with the running coupling generalization given later on. $\alpha d$ in the final term in \cref{eq:qgq}, is evaluated in \cref{app:a} as
\begin{equation}
  \alpha d=\frac{\alpha}{3\pi}\bigg(1+\frac{2M^{2}}{Q^{2}}\bigg)\sqrt{1-\frac{4M^{2}}{Q^{2}}}.
\label{eq:ad}
\end{equation}

The three terms on the right hand side of \cref{eq:qgq} are easy to understand. If there were no gluon branching, but only the possibility of decaying into the $c\bar{c}$ pair only the last term on the right hand side of \cref{eq:qgq} would be present and $G$ would decrease exponentially in $\ln Q$. This would be exactly as in the decay of unstable elements, with $\ln Q$ serving as a time. The first term on the right hand side of \cref{eq:qgq} simply says that if a $g\rightarrow g_{1}+g_{2}$ branching occurs the probability that no $c\bar{c}$ be produced is the probability that neither $g_{1}$ nor $g_{2}$ have a $c\bar{c}$ as part of their subsequent branchings. The second term on the right hand side of \cref{eq:qgq} is a probability conserving term associated with $g\rightarrow g_{1}+g_{2}$ branchings. \cref{eq:qgq} is identical in form to \cref{eq:c}, or \cref{eq:Gy}. of BMS.

\cref{eq:qgq} cannot be solved exactly. However, there are two limits where analytic solutions can be obtained, (i) $1-G\ll 1$ and (ii) $G\ll 1$.

When $1-G\ll 1$ it is convenient to introduce 
\begin{equation}
  G=1-P
\end{equation}
and to linearize the resulting equation for $P$. $P$ corresponds to the multiplicity of $c\bar{c}$ pairs in the decay. This gives
\begin{equation}
  Q\frac{\partial}{\partial Q}P(Q)=\frac{2\alpha N_{c}}{\pi}\int^{1}_{0}\frac{dx}{x}\big[P(xQ)+P((1-x)Q)-P(Q)\big]+\frac{\alpha}{3\pi}\Theta(Q-2M)
\label{eq:qpq}
\end{equation}
where we have simplified \cref{eq:ad} using the fact that, when $\alpha$ is small, decays of $g\rightarrow c\bar{c}$ are unlikely to occur near threshold. To the differential equation \cref{eq:qpq} we add the initial condition, $P(2M)=0$, reflecting the fact that there can be no $c\bar{c}$ production below threshold. \cref{eq:qpq} is easily solved by introducing
\begin{equation}
  \rho=\ln\frac{Q}{2M},\qquad \rho'=\ln\frac{xQ}{2M}
\end{equation}
giving
\begin{equation}
  \frac{\partial}{\partial \rho}P(\rho)=\frac{2\alpha N_{c}}{\pi}\int^{\rho}_{0}d\rho'\,P(\rho')+\frac{\alpha}{3\pi}\Theta(\rho),
\label{eq:rhop}
\end{equation}
where we have set $P((1-x)Q)-P(Q)$ equal to zero in our double logarithmic approximation, and where we now view $P$ as a function of the variable $\rho=\ln\frac{Q}{2M}$ rather than $Q$. Taking another derivative in $\rho$ gives
\begin{equation}
  \bigg(\frac{\partial^{2}}{\partial \rho^{2}}-\frac{2\alpha N_{c}}{\pi}\bigg)P(\rho)=\frac{\alpha}{3\pi}\delta(\rho).
\end{equation}
\cref{eq:rhop}, along with $P(0)=0$ and $P(\rho)=0$ for $\rho<0$, as well as $\partial P/\partial\rho=\alpha/3\pi$ at $\rho=0$, give
\begin{equation}
  P(\rho)=\frac{1}{6}\sqrt{\frac{\alpha}{2\pi N_{c}}}\Big(e^{\sqrt{\frac{2\alpha N_{c}}{\pi}}\rho}-e^{-\sqrt{\frac{2\alpha N_{c}}{\pi}}\rho}\Big)\Theta(\rho).
\label{eq:prho}
\end{equation}
Of course \cref{eq:prho} can only be used when $P\ll 1$,  the region where the linearization of \cref{eq:qgq} is valid. The growing exponential in \cref{eq:prho} corresponds to the growth of the total multiplicity of gluons~\cite{Mueller:1981ex,Bassetto:1982ma,khoze} in a jet decay in the leading double logarithmic, angular ordered, approximation. 

We can also solve \cref{eq:qgq} analytically in the  regime where $G\ll 1$. In this region of $\rho$ only the virtual term in \cref{eq:qgq} is important so 
\begin{equation}
  \frac{\partial}{\partial \rho}G(\rho)=-\frac{2\alpha N_{c}}{\pi}\int^{\rho}_{\rho_{0}}d\rho'\ G(\rho)
\label{eq:rhoG}
\end{equation}
where $\rho_{0}$ is chosen to be such that $G(\rho')$ is small when $\rho'>\rho_{0}$. This gives 
\begin{equation}
  G(\rho)=e^{-\frac{\alpha N_{c}}{\pi}(\rho-\rho_{0})^{2}}G(\rho_{0})
\label{eq:grho}
\end{equation}
identical to the form found by Levin and Tuchin~\cite{Levin:1999mw}. \cref{eq:grho} is valid when $G(\rho_{0})\ll 1$.

\subsection{Running coupling evaluation}
It is not hard to generalize our discussion to the case where running coupling effects are included. The running coupling is naturally put into \cref{eq:rhop} as 
\begin{equation}
  \frac{\partial}{\partial \rho}P(\rho)=\frac{2N_{c}}{\pi}\int^{\rho}_{0}d\rho'\,\alpha(\rho')P(\rho')+\frac{\alpha(\rho)}{3\pi}\Theta(\rho),
\label{eq:prunning}
\end{equation}
where the leading order form for $\alpha(\rho)$, 
\begin{equation}
  \alpha(\rho)=\frac{1}{2b(\rho+\bar{\rho})},\qquad \bar{\rho}=\ln\frac{2M}{\Lambda},
\end{equation}
will be used and with $\Lambda$ the usual QCD parameter and $b=(11N_{c}-2N_{f})/12\pi$. It is convenient to write \cref{eq:prunning} as
\begin{equation}
  \frac{\partial^{2}}{\partial \rho^{2}}P(\rho)-\frac{N_{c}}{\pi b(\rho+\bar{\rho})}P(\rho)=-\frac{\Theta(\rho)}{6\pi b(\rho+\bar{\rho})^{2}}+\frac{1}{6\pi b(\rho+\bar{\rho})}\delta(\rho)
\label{eq:prhod}
\end{equation}
which can be solved in terms of $I_{1}$ and $K_{1}$ Bessel functions. For $\rho/\bar{\rho}$ and $(\rho+\bar{\rho})$ large the solution to \cref{eq:prhod} takes the form~\cite{Munier:com}
\begin{equation}
  P(\rho)=C(\bar{\rho})\sqrt{(\rho+\bar{\rho})\frac{N_{c}}{\pi b}}I_{1}\bigg(\sqrt{2(\rho+\bar{\rho})\frac{N_{c}}{\pi b}}\bigg)
\end{equation}
where $C(\bar{\rho})$ is a constant, in $\rho$, given as an integral over $K_{1}$ and the terms on the right hand side of \cref{eq:prhod}. One again recognizes the $I_{1}$ function as giving the angular ordered, and running coupling, growth of the gluon multiplicity.

Similarly it is straightforward to put running coupling effects into \cref{eq:rhoG} which gives
\begin{equation}
  \frac{\partial}{\partial \rho} G(\rho)=-\frac{2N_{c}}{\pi}\int^{\rho}_{\rho_{0}}d\rho'\, \alpha(\rho') G(\rho)
\end{equation}
or 
\begin{equation}
  \frac{\partial}{\partial \rho}G(\rho)=-\frac{N_{c}}{\pi b}\ln\frac{\rho}{\rho_{0}}\, G(\rho).
\end{equation}
Thus
\begin{equation}
  G(\rho)=\exp\bigg[-\frac{N_{c}}{\pi b}\Big(\rho\ln\frac{\rho}{\rho_{0}}-(\rho-\rho_{0})\Big)\bigg]G(\rho_{0})
\end{equation}
replaces \cref{eq:grho} in the running coupling case.

\section{The origin of nonlinear evolution equations}
Now that we have seen how the nonlinear BMS equation can appear in evaluating certain properties of $c\bar{c}$ production in jet decay it is perhaps useful to review two other circumstances, besides non-global logarithms in jet decays, where conceptually identical nonlinear equations come up. After this review we shall attempt to give a more general picture when such nonlinear evolutions can be expected.

\subsection{Branching random walks and the FKPP equation\label{sec:branching}}
Our first example is from statistical physics and concerns properties of a branching random walk. Let us first review the phenomenon and then note the similarity with what we have just done in \cref{sec:prob}.

Consider a branching random walk of particles on the real $x$-axis starting from a single particle at $x=0$ at time $t=0$. A particle has a rate, $r_{12}$, to turn into two particles at the same $x$-value as the parent. A particle at $x$ also can carry out diffusion, moving to the left or right of $x$ over a time interval $dt$ according to $x\rightarrow x+\nu\sqrt{dt}$. $\nu$ is a gaussian random variable obeying
\begin{equation}
  \left<\nu\right>=0,\quad \frac{1}{2}\left<\nu^2\right>=x_{0}^{2},
\end{equation}
and where the diffusion occurs at a rate $r_{d}$. Suppose $Q(x,t)$ is the probability that, at time $t$, the rightmost particle in the branching random walk, starting from a single particle at $x=0$ when $t=0$, has not yet reached the value $x$, where $x>0$. It is simple to write an equation for $Q$ by taking a short time interval $dt$ and noting that, following the early branching or diffusions~\cite{Brunet,Munier:2014bba},
\begin{equation}
  Q(x,t+dt)=r_{12}dt\, Q^{2}(x,t)+r_{d}\left<Q(x-\nu\sqrt{dt},t)-Q(x,t)\right>+(1-r_{12}dt)Q(x,t).
\label{eq:qx}
\end{equation}
The first term on the right hand side of \cref{eq:qx} represents splitting times the product of the probabilities that neither of the daughter particles have descendents which reach $x$ by time $t$. The second term represents the requirement that none of the descendents of the original particle reach $x$ by time $t$ when the original particle diffuses over the time interval $dt$. The third term conserves probability and represents the probability that the parent does not branch during the time interval $0\leq t\leq dt$. To first order in $dt$, \cref{eq:qx} gives
\begin{equation}
  Q(x,t+dt)-Q(x,t)=dt\bigg[r_{12}Q^{2}+r_{d}x_{0}^{2}\frac{\partial^{2}}{\partial x^{2}}Q-r_{12}Q\bigg]
\end{equation}
or
\begin{equation}
  \frac{\partial Q}{\partial t}=r_{d}x_{0}^{2}\frac{\partial^{2}Q}{\partial x^{2}}-r_{12}Q+r_{12}Q^2.
\end{equation}
After rescaling $r_{12}t\rightarrow t$, $\frac{x}{x_{0}}\sqrt{\frac{r_{12}}{r_{d}}}\rightarrow x$ one gets
\begin{equation}
  \frac{\partial Q}{\partial t}=\frac{\partial^{2}Q}{\partial x^{2}}-Q+Q^{2},
\label{eq:fkpp}
\end{equation}
the FKPP equation. The initial condition for the process we have described above is~\cite{Brunet,Munier:2014bba}
\begin{equation}
  Q(x,0)=\Theta(x).
\label{eq:condition}
\end{equation}
At first sight \cref{eq:qgq,eq:fkpp} look very different. \cref{eq:qgq} is an equation in a single variable while \cref{eq:fkpp} has two variables. Nevertheless, they have much in common. They are both nonlinear equations having a stable fixed point at $G$, or $Q$, equal zero. \cref{eq:fkpp} has an unstable fixed point at $Q=1$ while \cref{eq:qgq} has a term $-\alpha dG$ driving the solution away from the $G=1$ fixed point of the remaining part of \cref{eq:qgq}. \cref{eq:qgq,eq:fkpp}, and similar equations which we shall review shortly, are used in similar ways in many physics applications. In the case of \cref{eq:qgq} one starts at the unstable fixed point at $G=1$ as the boundary condition and then evolution in $\rho$ drives $G$ to the $G=0$ stable fixed point. In the branching random walk one also takes an initial condition \cref{eq:condition} which for $x>0$ agrees with the unstable fixed point. Taking $Q=0$ for $x<0$ fits the stable fixed point. The interesting dynamics is in the motion of the stable fixed point solution, as a travelling wave, to take over the whole $x>0$ region. In both \cref{eq:qgq,eq:fkpp} it is the flow from $G$, or $Q$,$=1$ to $G$, or $Q$,$=0$ which is of interest.

Equations for $1-G$, and $1-Q$, both evolve away from zero in an exponential way, in $\rho$ in \cref{eq:prho} and in $t$ in the solution to \cref{eq:fkpp}. After a period of evolution $G$ goes to zero in a gaussian manner in $\rho$ while $Q$ goes to zero exponentially in $t$. The difference between the gaussian and exponential behavior is a detail. (The exact form of the $\rho$-dependence of $G$ will change when running coupling corrections are included but the fixed point, or near fixed point structure, will not change.

\subsection{The BK equation\label{sec:bk}}
Let us now turn to a widely used equation for high energy scattering, the Balitsky-Kovchegov (BK) equation. Consider the scattering of a quark-antiquark dipole of size $x_{01}=(\underline{x}_{1}-\underline{x}_{2})$ on a large nucleus. ($\underline{x}_{0}$ and $\underline{x}_{1}$ are the transverse coordinates of the quark and antiquark, respectively.) If $T$ is the scattering amplitude at rapidity $y$ the BK equation~\cite{Balitsky:1995ub,Kovchegov:1999yj} is usually written as
\begin{equation}
  \frac{\partial T(x_{01},y)}{\partial y}=\frac{\alpha N_{c}}{2\pi^{2}}\int d^{2}x_{2}\frac{x_{01}^{2}}{x_{02}^{2}x_{21}^{2}}\big[T(x_{02},y)+T(x_{21},y)-T(x_{01},y)-T(x_{02},y)T(x_{21},y)\big].
\label{eq:bk}
\end{equation}
For our purposes it is more useful to write \cref{eq:bk} as 
\begin{equation}
  \frac{\partial S(x_{01},y)}{\partial y}=\frac{\alpha N_{c}}{2\pi^{2}}\int d^{2}x_{2}\frac{x_{01}^{2}}{x_{02}^{2}x_{21}^{2}}\big[S(x_{02},y)S(x_{21},y)-S(x_{01},y)\big]
\label{eq:bkS}
\end{equation}
where
\begin{equation}
  S=1-T.
\end{equation}
The initial condition for \cref{eq:bk} or \cref{eq:bkS} is just the low energy amplitude $T(x_{01},0)$. Just like \cref{eq:fkpp} and similar to \cref{eq:qgq}, \cref{eq:bk,eq:bkS} are nonlinear equations with a stable fixed point at $S=0$ and an unstable fixed point at $S=1$. Starting near the unstable fixed point at zero, $T$ grows exponentially in $y$ just as $P(\rho)$ in \cref{eq:prho} grows exponentially. The approach to zero of $S$ as $y$ gets large is identical to the approach of $G$ to zero as given by \cref{eq:grho} with a change of $\rho$ to $y$.

One usually views \cref{eq:bk} as an equation which gives the BFKL growth to $T$ when $T$ is small and then imposes unitarity as $T$ gets near one. Our previous \cref{eq:qgq,eq:fkpp} are not based on unitarity but rather on the probability that something does not happen. \cref{eq:qgq} gives the probability that no $c\bar{c}$ pairs are produced while \cref{eq:fkpp} determines the probability that a one dimensional branching random walk starting at the origin at $t=0$ have no descendents to the right of $x$ at time $t$. The idea of unitarity \underline{seems} to be absent from these problems. On the other it does seem that we can interpret \cref{eq:bkS} in terms of something not happening. $|S(x_{01},y)|^{2}$ is the survival probability for the dipole to go through the nucleus. That is it is the probability that no inelastic collision with the nucleus occurs while the dipole passes through it. $S(x_{01},y)$ is the amplitude for no inelastic collision to occur. The evolution \cref{eq:bkS} just reflects the fact that when the dipole $x_{01}$ splits into two dipoles, $x_{02}$ and $x_{12}$, the amplitude that no interaction with the nucleus occur is the product of the amplitude that no interaction of the nucleus with dipole $x_{02}$ occur times the amplitude for no interaction with dipole $x_{12}$. Thus \cref{eq:bkS} has essentially the structure as \cref{eq:qgq}, the only difference being that (i) the initial condition for \cref{eq:qgq} is $G(Q)\big|_{2M}=0$ while the initial condition for \cref{eq:bkS} is the $S$-matrix at some (generally low) $y_{0}$ and (ii) the $\alpha dG$ term in \cref{eq:qgq} which allows $G(Q)\big|_{2M}=0$ to be a good initial condition since $G=0$ is not a fixed point of \cref{eq:qgq}.

\section{No $c\bar{c}$ production in dipole-nucleus collisions\label{sec:no}}
As a second example of a nonjet process where a BMS equation emerges we consider the probability, $G(x_{01},y)$, that no $c\bar{c}$ pairs be produced in a rapidity interval $\Delta y$ in the collision of a dipole of size $x_{01}$ on a large nucleus where the relative rapidity of the dipole and the nucleus is $y$. We specify the $\Delta y$ interval more precisely: Suppose a charm quark has energy $E$. Define $E_{0}$ by $\frac{2E_{0}}{M^{2}}=2R_{A}$ where $R_{A}$ is the nuclear radius and $M$ the charm quark mass. $E_{0}$ is then the minimum energy for the charm quark to have a coherence time as long as the nuclear diameter. Let $y_{0}$ be given by $E_{0}=M\cosh y_{0}$. The requirement on $G(x_{01},y)$ is that there be no charm quarks in the rapidity interval $y_{0}$ to $y_{0}+\Delta y$. We shall specify the size $\Delta y$ later on. According to the interpretation we gave \cref{eq:bkS} of $S$ representing the amplitude for no inelastic interaction of a dipole with a nucleus we no have almost the same equation for $G$. That is
\begin{align}
  \frac{\partial G(x_{01},y)}{\partial y}=\frac{\alpha N_{c}}{2\pi^{2}}\int d^{2}x_{2}\frac{x_{01}^{2}}{x_{02}^{2}x_{12}^{2}}&\big[G(x_{02},y)G(x_{12},y)-G(x_{01},y)\big]\nonumber\\
  &-\alpha^{2}N_{c}v G(x_{01},y)\Theta(y_{0}+\Delta y-y)\Theta(y-y_{0}),
\label{eq:bms}
\end{align}
where, as evaluated in \cref{app:b},
\begin{equation}
  \alpha^{2}N_{c}v=\frac{2\alpha^{2}N_{c}Q_{s}^{2}}{15\pi^{2}M^{2}}\ln(M^{2}x_{01}^{2})
\label{eq:Nv}
\end{equation}
with $Q_{s}^{2}$ the McLerran-Venugopalan saturation momentum of the nucleus. The initial condition for \cref{eq:bms} is $G(x_{01},y_{0})=1$. In principle there is no reason that the $\Delta y$ rapidity has to be of limited size and near, in rapidity, to the nucleus. However, in the more general case \cref{eq:Nv} will become more complicated and analytic solutions will not be possible. We assume $Q_{s}^{2}/M^{2}<1$ in getting \cref{eq:Nv}.

There are two issues that deserve comment. First, as we have noted before the BMS equation \cref{eq:bms} is not identical to \cref{eq:bkS} because of the $\alpha^{2}N_{c}vG$ term in \cref{eq:bms}. However, as we shall see below its behavior, both near $G=1$ and near $G=0$ is essentially identical to the solution of \cref{eq:bkS} for $S$. Secondly, $G$ is \cref{eq:bms} is a probability while $S$ is \cref{eq:bkS} is an amplitude. Nevertheless in each case they represent something not happening, no inelastic interaction for $S$ and no $c\bar{c}$ production for $G$. This we feel is the essential ingredient in getting a BMS or BK equation.

One may wonder why we have chosen a large nucleus both for \cref{eq:bkS,eq:bms}. The reason is that \cref{eq:bkS,eq:bms} properly treat fluctuations of the projectile dipole, $x_{01}$, but they do not correctly incorporate target fluctuations. In the case of a large nucleus one expects target fluctuations to be small but there is no good argument for that being the case for, say, a target proton. When \cref{eq:bkS} or \cref{eq:bms} are used with a proton target one is attempting to do a mean field evaluation where target fluctuations are neglected.

Now let us solve \cref{eq:bms} analytically first in the $G\simeq 1$ region and then in the $G\simeq 0$ region. When $G$ is near one it is convenient to define $P=1-G$ which obeys a linear equation, with initial condition $P(x_{01},0)=0$,
\begin{equation}
  \frac{\partial P(x_{01},y)}{\partial y}=\frac{\alpha N_{c}}{2\pi^{2}}\int\, d^{2}x_{2}\frac{x_{01}^{2}}{x_{12}^{2}x_{02}^{2}}\big[P(x_{12},y)+P(x_{02},y)-P(x_{01},y)\big]+c\ln(M^{2}x_{01}^{2})\Theta(\Delta y-y)\Theta(y)
\label{eq:py}
\end{equation}
where we have let $y\rightarrow y+y_{0}$ and with
\begin{equation}
  c=\frac{2\alpha^{2}N_{c}}{15\pi^{2}}\frac{Q_{s}^{2}}{M^{2}}
\label{eq:c}
\end{equation}
and where $(Q_{s}/2M)^{2}\ll 1$ has been assumed. \cref{eq:py} is solved in terms of the BFKL characteristic function $\chi(\lambda)$ as
\begin{equation}
  P(x_{01},y)=\int\frac{d\lambda}{2\pi i}\frac{c}{\lambda^{2}}\frac{e^{\lambda\ln (M^{2}x_{01}^{2})}}{\frac{2\alpha N_{c}}{\pi}\chi(\lambda)}
  \begin{cases}
    (e^{\frac{2\alpha N_{c}}{\pi}\chi(\lambda)y}-1)\Theta(\Delta y-y)\\
    e^{\frac{2\alpha N_{c}}{\pi}\chi(\lambda)y}(1-e^{-\frac{2\alpha N_{c}}{\pi}\chi(\lambda)\Delta y})\Theta(y-\Delta y)
  \end{cases}.
\label{eq:px}
\end{equation}
The $\lambda$-integral goes along $\mathrm{Re}\lambda=\frac{1}{2}$. Note that \cref{eq:px} gives $P(x_{01},y=0)=0$ and $\frac{\partial P(x_{01},y)}{\partial y}\big|_{y=0}=c\ln(M^{2}x_{01}^{2})$ as required by \cref{eq:py} and the condition that no $c\bar{c}$ pairs are produced in a low energy scattering. The BFKL growth of $P$ comes from taking the $\lambda=\frac{1}{2}$ saddle point, valid when $\frac{2\alpha N_{c}y}{\pi}/\ln(M^{2}x_{01}^{2})\gg 1$. One gets, for $(\alpha_{P}-1)\Delta y\ll 1$, and with $\alpha_{P}-1=\frac{2\alpha N_{c}}{\pi}\chi(\frac{1}{2})=\frac{\alpha N_{c}}{\pi}4\ln 2$,
\begin{equation}
  P(x_{01},y)=c\Delta y(M^{2}x_{01}^{2})^{1/2}\frac{e^{(\alpha_{P}-1)y}}{\sqrt{\frac{7}{2}\alpha N_{c}\zeta(3)y}}
\end{equation}
with a $y$-growth identical to that given by \cref{eq:bk} when $T$ is small.

When $G$ is small \cref{eq:bms} becomes
\begin{equation}
  \frac{\partial G}{\partial y}=-\frac{\alpha N_{c}}{2\pi^{2}}\int \frac{x_{01}^{2}d^{2}x_{2}}{x_{12}^{2}x_{02}^{2}}G(x_{01},y)
\label{eq:pGy}
\end{equation}
where the limits of integration in \cref{eq:pGy} are determined by $x_{12}^{2},x_{02}^{2}>1/Q_{s}^{2}(y)$ with $Q_{s}(y)$ the nuclear saturation momentum at rapidity $y$. Thus
\begin{equation}
  \frac{\partial G(x_{01},y)}{\partial y}=-\frac{\alpha N_{c}}{\pi}\ln\big[x_{01}^{2}Q_{s}^{2}(y)\big].
\label{eq:Gy}
\end{equation}
Using 
\begin{equation}
  Q_{s}^{2}(y)=Q_{s}^{2}(MV)e^{\frac{2\alpha N_{c}}{\pi}\frac{\chi(\lambda_{0})}{1-\lambda_{0}}y}
\label{eq:qs}
\end{equation}
where $\lambda_{0}$ is the usual saturation line eigenvalue satisfying $-\chi'(\lambda_{0})(1-\lambda_{0})=\chi(\lambda_{0})$. Using \cref{eq:qs} in \cref{eq:Gy} gives
\begin{equation}
  G(x_{01},y)=\exp\bigg[-\bigg(\frac{\alpha N_{c}}{\pi}\bigg)^{2}\frac{\chi(\lambda_{0})}{1-\lambda_{0}}(y-y_{0})^{2}\bigg]G(x_{01},y_{0})
\end{equation}
which is identical to the (fixed coupling) Levin-Tuchin result for the $y$-dependence of $S(x_{01},y)$ when $S$ is small.

Finally we note that in their original paper BMS initiated a study of energy flow, away from the various jet axes, in hadron-hadron collisions where two jets are produced. We believe there is much more to be done in hadron collisions and that there should be many observables where nonlinear equations arise. 
\begin{acknowledgments}
AM wishes to thank Stephane Munier for stimulating discussions. He also wishes to acknowledge support from the US Department of Energy.
\end{acknowledgments}

\appendix

\section{A derivation of \cref{eq:ad}\label{app:a}}
\begin{figure}[h]
  \centering
  \subfigure[]{\label{fig:dipole-gluon}
  \includegraphics[width=7cm]{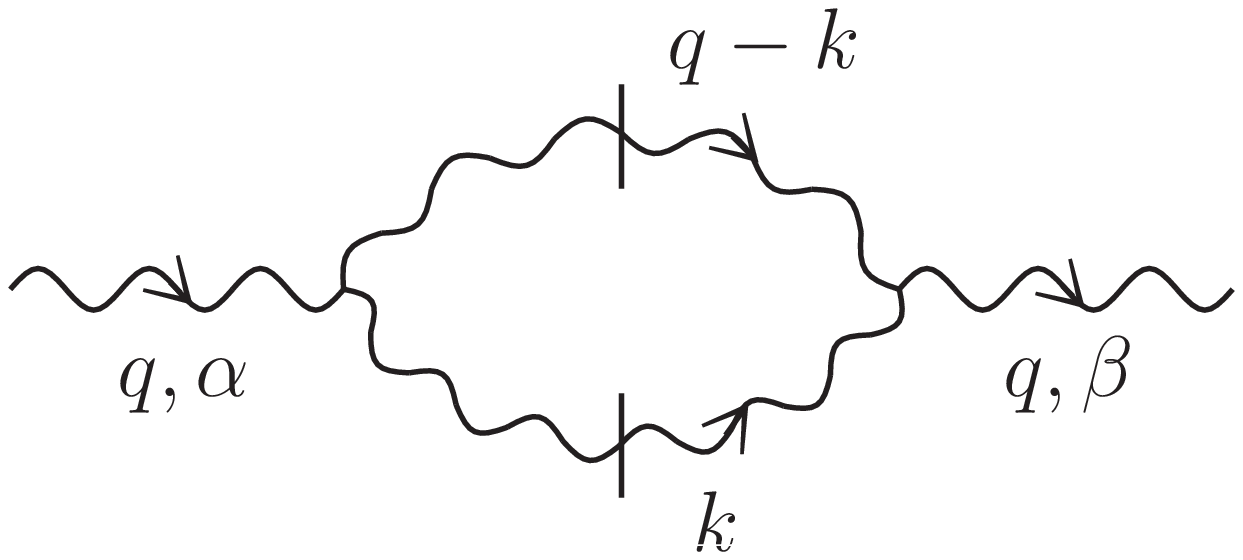}}
  \hspace{0.5in}
  \subfigure[]{\label{fig:dipole-quark}
  \includegraphics[width=7cm]{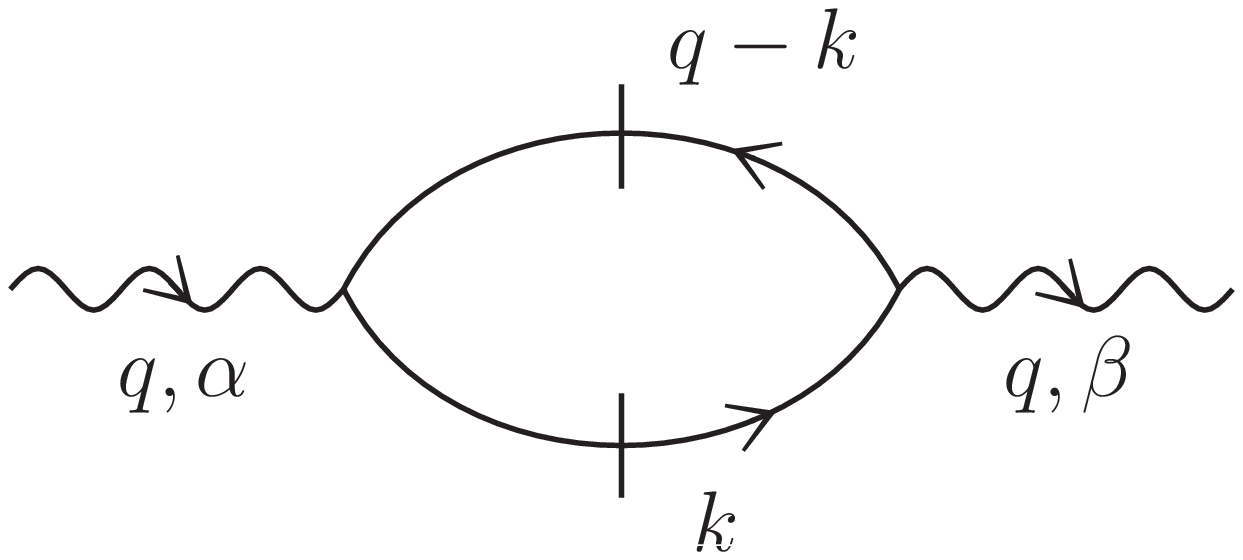}}
  \caption{}
\label{fig:dipole}
\end{figure}
In this appendix we calculate the gluon$\rightarrow c\bar{c}$ term, the final term on the right hand side of \cref{eq:qgq}, given by \cref{eq:ad}. The first term on the right hand side of \cref{eq:qgq} is the gluon splitting term which is well known. What we are going to calculate is the relative values of $g\rightarrow c\bar{c}$ compared to $g\rightarrow gg$ by evaluating the graphs given in \cref{fig:dipole} without regard to overall normalization. We begin with \cref{fig:dipole-gluon} which in the limit $k_{+}/q_{+}\ll 1$ is given in $A_{+}=0$ lightcone gauge as
\begin{equation}
  A_{\alpha\beta}=g_{\alpha\beta}g^{2}N_{c}(2q_{+})^{2}\int\frac{d^{4}k}{(2\pi)^{4}}2\pi\delta(k^{2})2\pi\delta((q-k)^{2})\frac{k_{\perp}^{2}}{k_{+}^{2}}
\label{eq:Aab}
\end{equation}
where the factors of $2q_{+}$ come from the eikonal vertices where the soft gluon $k$ hooks to the hard gluon $q$, and the $k_{\perp}^{2}/k_{+}^{2}$ factor comes from the polarization vector of the gluon $k$
\begin{equation}
  \epsilon_{\mu}^{\lambda}(k)=\big(\epsilon_{+}^{\lambda},\epsilon_{-}^\lambda,\epsilon_{\perp}^\lambda\big)=\bigg(0,\frac{\underline{\epsilon}^{\lambda}\cdot \underline{k}}{k_{+}},\underline{\epsilon}^{\lambda}\bigg).
\end{equation}
The $dk_{\perp}^{2}$ and $dk_{-}$ integrals are easily done in \cref{eq:Aab} to give
\begin{equation}
  A_{\alpha\beta}=g_{\alpha\beta}Q^{2}2\alpha N_{c}\int \frac{dk_{+}}{k_{+}}.
\label{eq:Aaa}
\end{equation}
The graph in \cref{fig:dipole-quark} is given by
\begin{equation}
  B_{\alpha\beta}=-\frac{1}{2}g^{2}\int d^{4}k\, 2\pi\delta(k^{2}-M^{2})2\pi\delta\big((q-k)^{2}-M^{2}\big)\mathrm{Tr}\big[\gamma_{\beta}(\gamma\cdot k+M)\gamma_{\alpha}\big(\gamma\cdot(q-k)-M\big)\big].
\end{equation}
One easily finds
\begin{equation}
  B_{\alpha\beta}=\frac{2\alpha}{3}\frac{(Q^{2}g_{\alpha\beta}-q_{\alpha}q_{\beta})}{2q_{+}}\bigg(1+\frac{2M^{2}}{Q^{2}}\bigg)\int^{k_{+}^{\max}}_{k_{+}^{\min}}dk_{+}
\end{equation}
where $k_{+}^{\max}$ and $k_{+}^{\min}$ are given by the maximum and minimum values of $k_{+}$ for which the integral $I$,
\begin{gather*}
  I=\int^{\infty}_{0}dk_{\perp}^{2}\, \delta\big((q-k)^{2}-M^{2}\big)\bigg|_{k_{-}=\frac{k_{\perp}^{2}}{2k_{+}}},
\end{gather*}
is not zero. One finds
\begin{equation}
  \bigg(\frac{k_{+}}{q_{+}}\bigg)_{\max,\min}=\frac{1}{2}\bigg[1\pm\sqrt{1-\frac{4M^{2}}{Q^{2}}}\bigg]
\end{equation}
so
\begin{equation}
  B_{\alpha\beta}=\frac{\alpha}{3}\bigg(1+\frac{2M^{2}}{Q^{2}}\bigg)\sqrt{1-\frac{4M^{2}}{Q^{2}}}\big(Q^{2}g_{\alpha\beta}-q_{\alpha}q_{\beta}\big).
\label{eq:Bab}
\end{equation}
The $q_{\alpha}q_{\beta}$ term does not contribute to $c\bar{c}$ production in jet decays so comparing the $g_{\alpha\beta}$ term in \cref{eq:Bab} with \cref{eq:Aaa} one finds 
\begin{equation}
  \alpha d=\frac{\alpha}{3\pi}\bigg(1+\frac{2M^{2}}{Q^{2}}\bigg)\sqrt{1-\frac{4M^{2}}{Q^{2}}}
\end{equation}

\section{Evaluation of \cref{eq:Nv}\label{app:b}}
\begin{figure}[h]
  \centering
  \includegraphics[width=6cm]{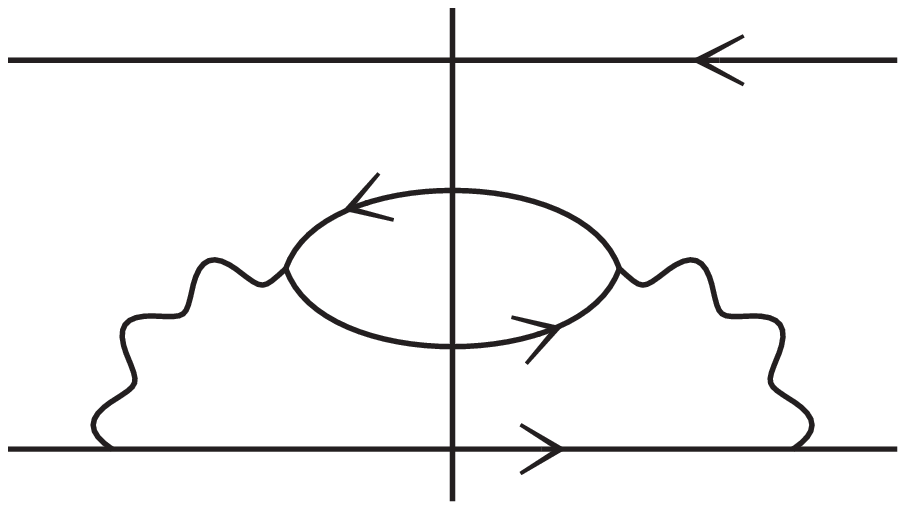}
\caption{}
\label{fig:dipole-radiation}
\end{figure}

The last term on the right hand side of \cref{eq:bms} comes from the graph shown in \cref{fig:dipole-radiation} and is given in \cref{eq:Nv} in the limit $Q^{2}_{s}x_{01}^{2}\gg 1$, $Q_{s}^{2}/M^{2}\ll 1$, with $Q_{s}^{2}$ the McLerran-Venugopalan saturation momentum. It is not difficult to directly evaluate the graph of \cref{fig:dipole-radiation}, and we have done this calculation, however it can also be obtained in a simple limit of a much more elaborate calculation given by Kovchegov and Tuchin (KT)~\cite{Kovchegov:2006qn}. We now give a set of steps to get from KT to $\alpha^{2}N_{c}v$ in \cref{eq:bms,eq:Nv}.

(i) Start with Eq.~(17) of KT which is a formula for inclusive quark or antiquark production in a proton-nucleus collision. Integrate Eq.~(17) of KT over $d^{2}k$ to get the total quark or antiquark production and divide by 2 to get the pair production. The identification with $\alpha^{2}N_{c}v$ is
\begin{equation}
  \alpha^{2}N_{c}v=\frac{1}{4}\frac{1}{(2\pi)^{2}}\int d^{2}x_{1}d^{2}x_{2}\int^{1}_{0}d\alpha\, \sum^{3}_{i,j=1}\Phi_{ij}(\underline{x}_{1},\underline{x}_{2};\underline{x}_{1},\underline{x}_{2},\alpha)\Xi_{ij}(\underline{x}_{1},\underline{x}_{2},\underline{x}_{1},\underline{x}_{2},\alpha)
\label{eq:aNv}
\end{equation}
with the notation as in KT.

(ii) In Eq.~(14) of KT identify $\frac{1}{4}x_{\perp}^{2}\ln(\frac{1}{x_{\perp}\Lambda})Q_{s}^{2}$ with what we have called $\frac{1}{8}x_{\perp}^{2}Q_{s}^{2}(\mathrm{MV})$ in this paper.

(ii) To evaluate the $\Phi_{ij}$ from Eq.~(12) of KT we must first evaluate $F_{0}$, $F_{1}$ $F_{2}$ from Eqs.~(6) (7) and (8) of KT. Since the heavy quark is the hardest scale in our problem take $M^{2}\gg 1/u^{2}$, $q^{2}$ in Eqs.~(6)-(8) of KT. This gives $F_{0}=0$, $F_{1}=K_{0}(Mx_{12})/u$, $F_{2}=K_{1}(Mx_{12})M/u$. 

(iv) Use the above $F_{i}$ in Eq.~(12) of KT to get 
\begin{equation}
  \Phi_{11}=\Phi_{22}=-\Phi_{12}=4\bigg(\frac{\alpha}{\pi}\bigg)^{2}C_{F}\frac{M^{2}}{u^{2}}\big[K_{0}^{2}(Mx_{12})+[\alpha^{2}+(1-\alpha)^{2}]K_{1}^{2}(Mx_{12})\big],
\label{eq:phi}
\end{equation}
with the $\Phi_{i3}=0,\, i=1,2,3$.

(vi) Expand the exponentials in Eq.~(14) of KT to first order in $Q_{s}^{2}$.

(vii) Using \cref{eq:phi} and the first order expansion of the $\Xi$ in \cref{eq:aNv} gives \cref{eq:Nv}.


\begin{thebibliography}{99}
\bibitem{Banfi:2002hw} 
  A.~Banfi, G.~Marchesini and G.~Smye,
  JHEP {\bf 0208}, 006 (2002)
  [hep-ph/0206076].

\bibitem{Balitsky:1995ub} 
  I.~Balitsky,
  Nucl.\ Phys.\ B {\bf 463}, 99 (1996)
  [hep-ph/9509348].

\bibitem{Kovchegov:1999yj}
  Y.~V.~Kovchegov,
  Phys.\ Rev.\ D {\bf 60} (1999) 034008
  [hep-ph/9901281].
  Phys.\ Rev.\ D {\bf 61}, 074018 (2000)
  [hep-ph/9905214].

\bibitem{Dasgupta:2001sh} 
  M.~Dasgupta and G.~P.~Salam,
  Phys.\ Lett.\ B {\bf 512}, 323 (2001)
  [hep-ph/0104277].

\bibitem{Kuraev:1977fs} 
  E.~A.~Kuraev, L.~N.~Lipatov and V.~S.~Fadin,
  Sov.\ Phys.\ JETP {\bf 45}, 199 (1977)
  [Zh.\ Eksp.\ Teor.\ Fiz.\  {\bf 72}, 377 (1977)].

\bibitem{Balitsky:1978ic} 
  I.~I.~Balitsky and L.~N.~Lipatov,
  Sov.\ J.\ Nucl.\ Phys.\  {\bf 28}, 822 (1978)
  [Yad.\ Fiz.\  {\bf 28}, 1597 (1978)].

\bibitem{Marchesini:2003nh} 
  G.~Marchesini and A.~H.~Mueller,
  Phys.\ Lett.\ B {\bf 575}, 37 (2003)

\bibitem{Marchesini:2004ne} 
  G.~Marchesini and E.~Onofri,
  JHEP {\bf 0407}, 031 (2004)
  [hep-ph/0404242].

\bibitem{Hatta:2008st} 
  Y.~Hatta,
  JHEP {\bf 0811}, 057 (2008)
  [arXiv:0810.0889 [hep-ph]].

\bibitem{Caron-Huot:2015bja} 
  S.~Caron-Huot,
  arXiv:1501.03754 [hep-ph].

\bibitem{Neill:2015nya} 
  D.~Neill,
  arXiv:1508.07568 [hep-ph].

\bibitem{Mueller:1981ex} 
  A.~H.~Mueller,
  Phys.\ Lett.\ B {\bf 104}, 161 (1981).

\bibitem{Ermolaev:1981cm} 
  B.~I.~Ermolaev and V.~S.~Fadin,
  JETP Lett.\  {\bf 33}, 269 (1981)
  [Pisma Zh.\ Eksp.\ Teor.\ Fiz.\  {\bf 33}, 285 (1981)].

\bibitem{Bassetto:1982ma} 
  A.~Bassetto, M.~Ciafaloni, G.~Marchesini and A.~H.~Mueller,
  Nucl.\ Phys.\ B {\bf 207}, 189 (1982).

\bibitem{khoze} 
  Yu.L. Dokshitzer, V.A. Khoze, A.H. Mueller, and S.I. Troyan, \textit{Basics of Perturbative QCD} (Editions Frontieres, Gif-sur-Yvette, 1991).

\bibitem{Levin:1999mw} 
  E.~Levin and K.~Tuchin,
  Nucl.\ Phys.\ B {\bf 573}, 833 (2000)
  [hep-ph/9908317].
  Nucl.\ Phys.\ A {\bf 691} (2001) 779
  [hep-ph/0012167].

\bibitem{Iancu:2003uh} 
  E.~Iancu and A.~H.~Mueller,
  Nucl.\ Phys.\ A {\bf 730}, 460 (2004)
  [hep-ph/0308315].

\bibitem{McKean}
H.~P.~McKean, Comm.\ Pure.\ Appl.\ Math.\ {\bf 28} (1975) 323

\bibitem{Bramson}
M.~D.~Bramson, Mem.\ Amer.\ Math.\ Soc. 44 (1983) 285

\bibitem{Brunet}
E.~Brunet, B.~Derrida,\ EPL\ 87,\ (2009)\ 60010

\bibitem{Munier:2014bba} 
  S.~Munier,
  Sci.\ China Phys.\ Mech.\ Astron.\  {\bf 58}, no. 8, 81001 (2015)
  [arXiv:1410.6478 [hep-ph]].

\bibitem{Fisher}
R.~A.~Fisher, Ann. Eugenics 7 (1937) 355

\bibitem{Kolmogorov}
A.~Kolmogorov, I.~Petrovsky, and N.~Piscounov, Moscow\ Univ.\ Bull.\ Math.\ A1,\ (1937) 1 

\bibitem{Munier:2003sj} 
  S.~Munier and R.~B.~Peschanski,
  Phys.\ Rev.\ D {\bf 69}, 034008 (2004)
  [hep-ph/0310357].

\bibitem{Munier:com}
  Stephane Munier, private communication.

\bibitem{Iancu:2004es} 
  E.~Iancu, A.~H.~Mueller and S.~Munier,
  Phys.\ Lett.\ B {\bf 606}, 342 (2005)
  [hep-ph/0410018].

\bibitem{Brunet:2006}
E.~Brunet, B.~Derrida, A.~H.~Mueller and S.~Munier, Phys.\ Rev.\ E\ {\bf 73},\ 056126\ (2006)

\bibitem{Mueller:2014fba} 
  A.~H.~Mueller and S.~Munier,
  Phys.\ Lett.\ B {\bf 737}, 303 (2014)
  [arXiv:1405.3131 [hep-ph]].

\bibitem{Mueller:1993rr} 
  A.~H.~Mueller,
  Nucl.\ Phys.\ B {\bf 415}, 373 (1994).

\bibitem{Kovchegov:2012mbw} 
  Y.~V.~Kovchegov and E.~Levin, ``Quantum chromodynamics at high energy''. Cambridge Monographs on Particle Physics, Nuclear Physics and Cosmology, Cambridge, UK (2012)

\bibitem{Kovchegov:2006qn} 
  Y.~V.~Kovchegov and K.~Tuchin,
  Phys.\ Rev.\ D {\bf 74}, 054014 (2006)
  [hep-ph/0603055].
\end{thebibliography}
\end{document}